# Application of Wavelet-based Active Power Filter in Accelerator Magnet Power Supply

Guo Xiaoling（郭晓玲）, Cheng Jian（程建）

Institute of High Energy Physics, Chinese Academy of Sciences, Beijing 100049, China

**Abstract:** As modern accelerators demand excellent stability to magnet power supply (PS), it is necessary to decrease harmonic currents passing magnets. Aim at depressing rappel current from PS in Beijing electron-positron collider II, a wavelet-based active power filter (APF) is proposed in this paper. APF is an effective device to improve the quality of currents. As a countermeasure to these harmonic currents, the APF circuit generates a harmonic current, countervailing harmonic current from PS. An active power filter based on wavelet transform is proposed in this paper. Discrete wavelet transform is used to analyze the harmonic components in supply current, and active power filter circuit works according to the analysis results. At end of this paper, the simulation and experiment results are given to prove the effect of the mentioned Active power filter.

## 1  Introduction

Most magnet power supplies of accelerator in BEPC II are high-precision direct current sources. The quality of current is one of the important factors affecting the stability of beam orbit, and therefore, how to eliminate the current harmonics is a hot spot in power supply filed.

Using LC passive filter to decrease rappel current is a common method to improve PS's quality, and installing active power filter (APF) is another way. Usually, the cost in first scheme is very high although it can evidently reduce harmonic current. Furthermore, as the supply's power increases, the size of filter increases, and its cost becomes much higher. The second way of APF has an advantage of low cost[1].

Fig.1 presents the principle diagram of APF. APF is composed of two parts, which are arithmetic logical unit (ALU) and compensation circuit. Between them, the ALU circuit is to detect the rappel current form PS to get the information of harmonics, and compensation circuit is to generate a compensating current with same value but opposite direction, offsetting the harmonic from the PS[1].

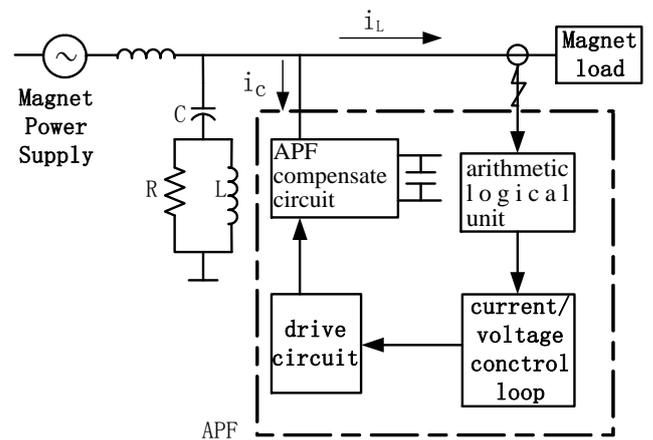

Fig. 1.  Principle diagram of APF.

Beyond any doubt that APF can eliminate harmonic current. However, it is difficult to get the values of harmonic's amplitude and phase, causing difficulty to control the APF circuit. This is one of the main reasons for limiting the development and application of APF technology. Wavelet decomposition algorithm provides a new approach to harmonic analysis.

## 2  Wavelet analysis[2]

Wavelet analysis is considered to be a breakthrough of Fourier analysis. It provides an adjustable window between time and frequency. The window becomes narrow automatically when observing the high-frequency signals, and it widens when focusing the low-frequency signals, namely it has the characteristic of zoom in, which make it very suitable for the harmonic analysis. The following is introducing wavelet analysis.

If $\psi \in L^2(R)$ satisfies the admissibility condition of $\int_R \psi(t)dt=0$, then $\psi$ can be used as a mother wavelet function, and then the continuous wavelet transform (CWT) to signal x(t) can be defined as follow.

$$WT_x(a,b) = a^{1/2} \int_{-\infty}^{\infty} x(t)\overline{\psi}\left(\frac{t-b}{a}\right)dt = <x(t), \psi_{a,b}(t)> .$$
(1)

Where $a, b \in R$, $a > 0$ is the scaling parameter, and b is the corresponding shifting parameter, $\psi_{a,b}(t)$ is the is the scaling and shifting of $\psi(t)$.

Discrete wavelet transform (DWT) makes it possible to realize WT in digital processing. It divides the original signal into separated frequency bands, so it is possible to analyze higher frequency components for each band independently. The input at each stage is always split into two bands, and then the higher band becomes one of the outputs, while the lower band is further split into two bands. This procedure is continued till a desire resolution is achieved, from which the recursion Equations of decomposing coefficient are shown as follows:

$$\begin{cases} a_{j+1}(k) = \sum_m h_0(m-2k)a_j(m) \\ d_{j+1}(k) = \sum_m h_1(m-2k)a_j(m) \end{cases}$$
(2)

Where $h_0(k)$ refers to unit sampling response of low-pass digital filter, and $h_1(k)$ refers to unit sampling response of high-pass digital filter, respecting $a_j$ and $d_j$ are the detail parts and the approaching parts of the analysis results. Making j increase from 0 gradually, then the signals in different bandwidth can be obtained as shown in fig.2.

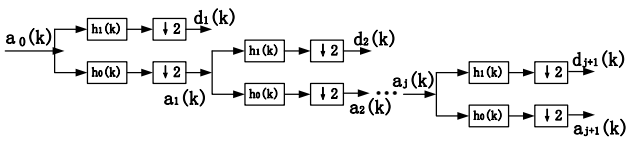

Fig. 2. Process of decomposing using wavelet analysis.

The data can be decomposed accommodate the process shown in Fig.2, but the decomposed result data cannot directly display the ripple current, they need to be reconstructed as shown in Fig.3.

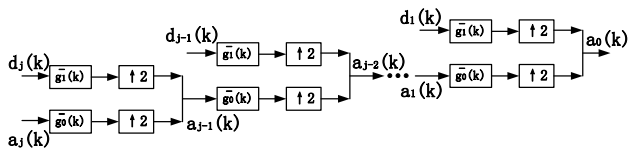

Fig. 3. Process of wavelet reconstructing.

## 3 Wavelet-based harmonic analysis

For the ripple currents from parts of BEPC II's magnet power supplies are not minute enough, a wavelet-based APF is designed based on the mentioned theory.

The APF referred to above adopts full bridge inverter circuit as its main circuit, and analyze the sampled data using wavelet transform. The circuit uses the classic PI control in its current loop, which is common in power supply, so it isn't described in this article. However, harmonic analysis is an important part in APF technology, and using wavelet is novelty, so it will be discussed in detail below.

### 3.1 Sampling frequency

As load varies, the concerned bandwidth is different. The bandwidth cared about is called interesting-bandwidth. Set the sampling frequency is $f_{sample}$, and sampled data is x(n), according to Whittaker-Shannon sampling theorem, then the bandwidth of x(n) is $f_{sample}/2$. Obviously, the sampling frequency should be at least 2 times interesting-bandwidth. To reduce the amount of calculation, make sampling frequency to be 2 times interesting-bandwidth, so that the bandwidth of x(n) is exactly interesting-bandwidth. As to accelerator magnet power supply, harmonics below 1000Hz need to be reduced, so it is appropriate to set sampling rate to 2048Hz.

### 3.2 Wavelet function

The selection of mother wavelet is not only a practical difficulty, but also a key point. Because there is no standard or reference, and what's more it will directly affect the result of analyzing. Matlab provide an effective way to choose mother wavelet, which is simulation. Db5 wavelet basis function is selected to be mother wavelet according to simulation result, but the simulation results are not shown in this paper due to space limitation.

### 3.3 Calculations[3]

As the sample-rate and mother-wavelet is determined, the current data can be analyzed. For db5 wavelet function, the coefficients are as follow:

$h_0$=[0.00235871396953395;-0.00889593505097710;-0.00441340005417915;0.054851329321067;-0.022800565941773 5;-0.171328357691468;0.0978834806739039;0.512163472129598;0.426971771352514;0.113209491291779].

$h_1$=[-0.113209491291779;0.426971771352514;-0.512163472129598;0.0978834806739039;0.171328357691468;-0.0228005659417735;-0.0548513293210670;-0.004413400054179

15;0.00889593505097710;0.00235871396953395].
$g_0$=[0.2264189825835584;0.8539435427050283;1.0243269442591967;0.1957669613478078; -0.3426567153829353;-0.0456011318835469;0.1097026586421339;-0.0088268001083583;-0.0177918701019542;0.0047174279390679].
$g_1$=[0.00471742793906790;0.0177918701019542;-0.00882680010835830;-0.109702658642134; -0.0456011318835469;0.342656715382935;0.195766961347808;-1.02432694425920;0.853943542705028;-0.226418982583558].

Set $a_0=x(n)$, then the data can be analysis according to Eq. (2). Each decomposetion divides the data bandwidth into 2 parts averagely, so the DC value is distinguished from original data after 5 times of analysis.

## 4 simulation and experiment

To demonstrate the validity of this wavelet analysis and the effect of this APF, the work process of the mentioned APF is simulated by software of Matlab, and then the experiment results are also be shown to verify the Matlab simulation results.

### 4.1 Matlab simulation

Most accelerator magnet power supplies outputs dc, so the simulation modeled on a three-phase bridge rectifier of stabilized current supply. Set the outputted dc current ($I_{p-p}$) is 100A, and the harmonics' amplitude of 50 Hz, 100 Hz, 150 Hz and 300 Hz are respectively 10mA, 10mA, 10mA and 70mA, which is shown in Fig.4.

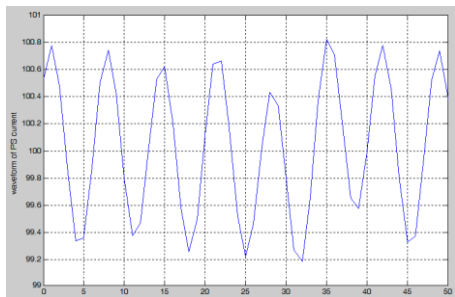

Fig. 4.    Waveform of PS current.

Obtained spectrum bandwidth and waveform after decomposition are shown in Fig.5, in which $a_j$ denotes low frequency signals while $d_j$ shows the high frequency ones.

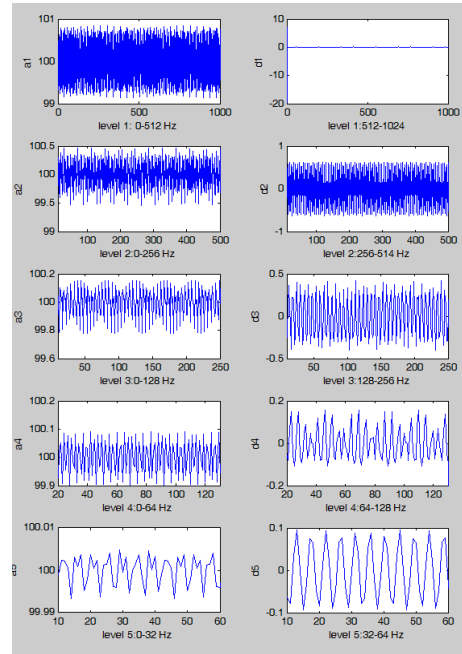

Fig. 5.    Waveform of wavelet-decomposition process simulation.

Reconstruction the harmonics within 1024 Hz into a signal named harmonic-refactoring component, then control the APF circuit to output current with same amplitude but opposite direction, and the outputted current is shown in Fig. 6. After installing this APF to magnet, the current through magnet is optimized as displayed in Fig.7.

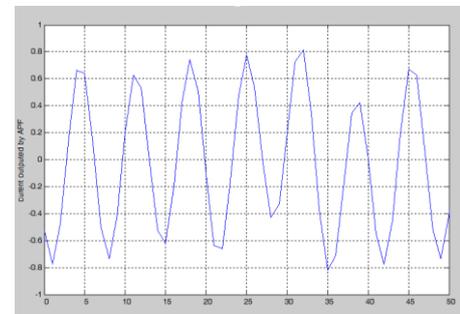

Fig. 6.    Current waveform of APF circuit.

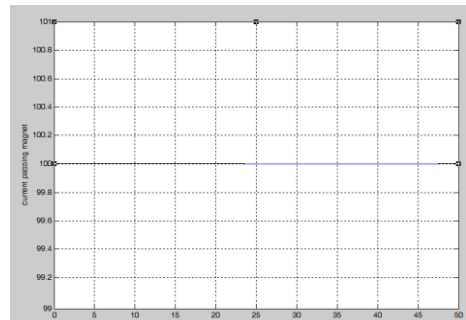

Fig. 7.    Final current waveform through load.

### 4.2 prototype experiment

The experiment is conducted on a 5A prototype PS,

whose voltage is 10V. because the current waveform is hardly displayed by oscilloscope, a resistance is chosen as load, so that the status of voltage is exactly the same with the current. The power supply's output voltage is displayed in Fig.8, and Fig.9 shows the output voltage of APF.

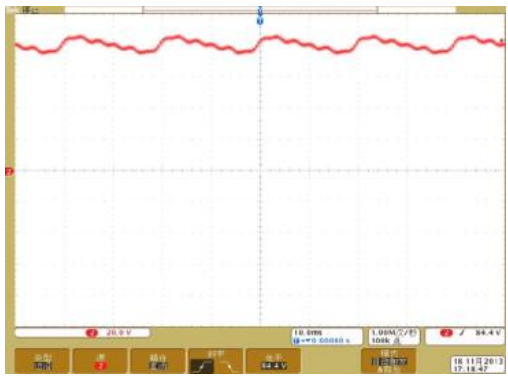

Fig. 8.　Waveform of PS output voltage.

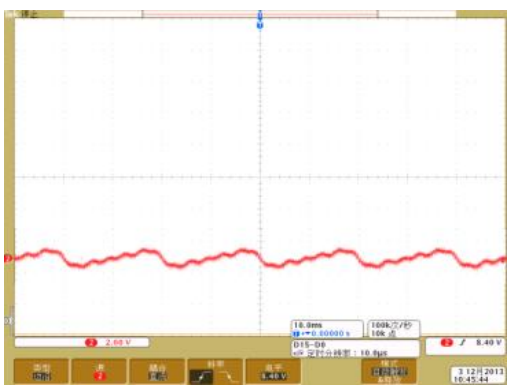

Fig. 9.　Waveform of APF output voltage.

The power supply and APF connect the resistance in parallel, and the final waveform of voltage across load is shown in Fig.10.

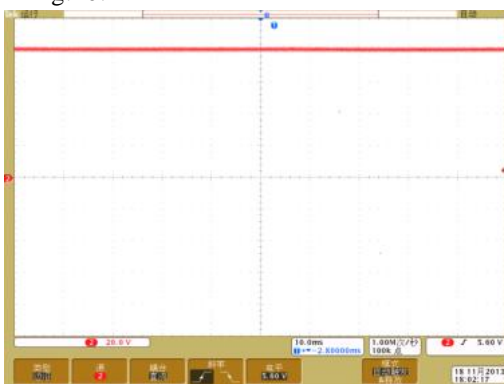

Fig. 10.　Waveform of voltage across the resistance

## 5 Conclusion

The wavelet-based APF mentioned in this paper can decrease rappel current effectively. By using wavelet analysis, the APF aimed bandwidth can be selected and modified easily. The simulation reveals its work process, and the experiment result approves the description in this paper, and also agrees the outcome of Matlab simulation.

## Refferences